\documentclass[a4paper,10pt,prl,twocolumn]{revtex4}
\usepackage[dvips]{graphics,color}
\usepackage{amsmath}
\usepackage{amsfonts}

\usepackage{amssymb}
\usepackage{amstext}
\usepackage[sort&compress]{natbib}
\usepackage{graphics,graphicx}
\usepackage{psfrag}
\newcommand{\ignore}[1]{}

\newcommand{\be}{\begin{equation}}
\newcommand{\ee}{\end{equation}}
\newcommand{\bea}{\begin{eqnarray}}
\newcommand{\eea}{\end{eqnarray}}

\newcommand{\up}{\uparrow}
\newcommand{\down}{\downarrow}

\newcommand{\ket}[1]{\left\vert #1    \right\rangle }
\newcommand{\bra}[1]{\left\langle   #1  \right\vert}

\begin{document}

\title{Double wells, scalar fields and quantum phase transitions
in ions traps}

\author{A. Retzker}
\author{R. Thompson}
\author{D. Segal}
\author{M.B. Plenio} \affiliation{Institute for Mathematical
Sciences, Imperial College London, SW7 2PE, UK} \affiliation{QOLS,
The Blackett Laboratory, Imperial College London, Prince Consort
Rd., SW7 2BW, UK}

\date{\today}

\maketitle

{\bf
Since Hund's work on the ammonia molecule\cite{hund1927},
the double well potential has formed a key paradigm
in physics. Its importance is further underlined by the
central role it plays in the Landau theory of phase
transitions\cite{LandauTheory}. Recently, the study of
entanglement properties of many-body systems has added
a new angle to the study of quantum phase transitions of
discrete and continuous degrees of freedom, i.e., spin
\cite{Osterloh 02,Jin 04} and harmonic chains
\cite{Audenaert EPW 02,Cramer EP 07}.
Here we show that control of the radial degree of freedom
of trapped ion chains allows for the simulation of linear
and non-linear Klein-Gordon fields on a lattice, in
which the parameters of the lattice, the non-linearity
and mass can be controlled at will. The system may be
driven through a phase transition creating a double
well potential between different configurations of the
ion crystal. The dynamics of the system are controllable,
local properties are measurable and tunnelling in the
double well potential would be observable.
}

The development of ion trap technology enables
precise control of the internal and external degrees
of freedom in strings of ions \cite{Leibfried2003}.
This  motivated proposals for the realization of
massive scalar fields in ion traps \cite{retzker2005,Alsing2005}.
These suggestions employed the longitudinal degrees of
freedom only, leading to significant restrictions both
in the accessible field theories, their mass etc, as
well as the extraction of local properties of the system.
We note that critical behavior is not experimentally accessible
in such systems. More recently the use of radial modes
in ion traps for the simulation of spin systems
\cite{Porras C 04,Porras2005a} and Bose-Einstein condensation
of phonons \cite{Porras2004a} has been proposed.

\begin{figure}
\begin{center}
\psfrag{a}{$\omega_x>\omega_c$}\psfrag{b}{$\omega_x<\omega_c$}\psfrag{c}{$\omega_x=\omega_c$}
\includegraphics[width=0.47\textwidth,height=0.15\textheight]{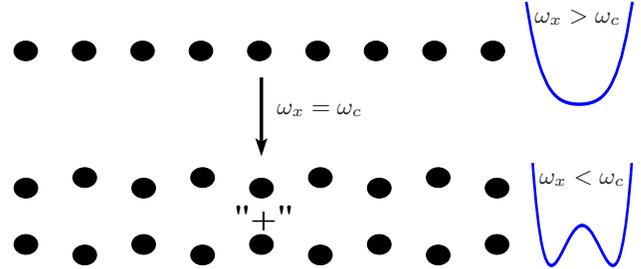}
\end{center}
\caption{The geometric transition between a linear formation and a
2D zigzag potential. The rotational symmetry has to be broken by the
electrode structure. By controlling the potential a double well
structure could be realized for which the ground state is a
superposition of the two zig zag configurations. }\label{trans}
\end{figure}

Here we suggest the use of the radial degrees of freedom
of the ions for the creation of a scalar field whose mass
we may adjust freely by varying the radially confining
potential. Crucially, in this way we have access to non-linear
fields if we allow the system to approach its configurational
phase transition where it changes from a linear to a zig-zag
structure. While this transition is well known at the
classical, high-temperature, level,
\cite{Dubin1999,Rafac1991,Wineland1987,Birkl KW 92,Casdorff1998}
at zero temperature it realizes a quantum phase transition
in the quantum field describing the system. Furthermore, at
zero temperature
the groundstate of the zig-zag structure is degenerate
(see Fig.\ref{trans}) and thus realizes a double well
structure. Parameters such as the width and depth of
the double well depend on the distance from the phase
transition point and may thus be controlled precisely
thanks to the high degree of control within reach in ion
trap experiments. This approach compares favorably to the
artificial creation of an electromagnetic double-well
potential for a single ion by applied electric fields,
as this would require field gradients and a degree of
control that are not likely to be accessible in the
foreseeable future.
The crucial aspect of our proposal is the fact that the
presence of several ions in the string leads to the
automatic creation of a well controlled double well
potential.

We will demonstrate that the above QPT may be characterized by the
potential seen by a single normal mode, namely the mode of lowest
frequency.
Then the realization of a QPT requires the ability to experimentally
adjust the potential perceived by the normal mode,
\be
        V(x) = V(0) + \frac{1}{2} a(\omega)x^2 +
        \frac{1}{4}b(\omega)x^4 + \frac{1}{6}c(\omega)x^6 + \ldots ,
        \label{landau}
\ee where $x$ is the degree of freedom of the `important' normal
mode and $\omega$ is the external trap potential frequency analogous
to the control parameter in the Landau theory. A second order
transition occurs when $a(\omega)=0$ and $b(\omega)>0$. A First order
transition occurs when the symmetry is
broken by a cubic term.

In the remainder of this paper we first discuss the entanglement
structure of the system in both the linear and the non-linear
regimes, i.e. next-to or away from the phase transition. Then
we present the non-linear Hamiltonian that is accessible
experimentally and show how it enables the creation and full
control of a double well potential. Finally, we discuss carefully
possible measurement schemes to verify our predictions.

%
{\em The Hamiltonian --}
In this work we consider chains of ions of mass $m$ that are
subject to external confining potentials in the radial (x-y)
plane and axial (z) direction as well as their mutual electrostatic
repulsion. The external potentials are harmonic and we will
assume that the confining potential in the y-direction is much
stronger than in the x-direction, resulting in an essentially
two-dimensional problem. This may be achieved both in linear
ion traps and Penning traps \cite{axialization}.
Thus the Hamiltonian for $N$ ions is of the form
\begin{eqnarray}
        H &=& \sum_{i=1}^{N} \left[\frac{{\hat p}_i^2}{2m} +
        \frac{1}{2} m\omega_x^2 {\hat x}_i^2 +
        \frac{1}{2} m \omega_z^2{\hat z}_i^2
        \right]\nonumber\\
        &&+ \frac{e^2}{4\pi\epsilon_0}\sum_{i > j}^N
        \frac{1}{\sqrt{({\hat x}_i-{\hat x}_j)^2
        + ({\hat z}_i-{\hat z}_j)^2}}
        \label{Hamiltonbasic}
\end{eqnarray}
where ${\hat x}_i$ is the distance of the i-th ion from the axis,
${\hat z}_i$ is the position on the axis and we have fixed the
strength of the confining potential in the axial direction.
All the numerical results in this work assume $\omega_z=1MHz$.
One may now determine the equilibrium arrangement of the electrostatic
problem posed here. There is a critical value $\omega_c$
($\omega_c\approx 3N\omega_z/(4\sqrt{\log N})$(for $N\gg
1$)\cite{Morigi F 04}) of the external confining potentials
$\omega_x$ for which we observe a transition from a linear chain
($\omega_x \ge \omega_c$) to a zig-zag configuration ($\omega_x \le
\omega_c$) as shown in fig. \ref{trans}. In the former case the
equilibrium positions in the axial direction do not depend on
$\omega_x$ while they become functions of $\omega_x$ in the latter.
We will be interested in the regime of small displacements ${\hat
x}_i$ when the coupling between radial and axial modes may be
neglected \cite{nonlinear}. Assuming the system is cooled close to
its ground state, we may then replace the operators ${\hat z}_i$ by
their equilibrium positions which we denote by $z_i$. Then we perform
a fourth order Taylor expansion of the potential to obtain the
effective Hamiltonian
\bea
        H &=& \frac{1}{2 m} \sum_{i=1}^N {\hat p}_i^2
        + \frac{1}{2}m\sum_{ij=1}^N \gamma_{ij} {\hat x}_i
        {\hat x}_j + \sum_{i=1}^N b_i {\hat x}_i^4 \nonumber\\
        &+& \sum_{ij=1}^N \alpha_{ij} {\hat x}_i^2 {\hat x}_j^2
        + \kappa_{ij} {\hat x}_i^3 {\hat x}_j,
        \label{Hamiltonexpanded}
\eea where
$\gamma_{ii}=\omega_x^2-\sum_j\frac{e^2}{2m\pi\epsilon_0\vert
z_i-z_j\vert^3}$, $b_i =\frac{1}{4!}
\sum_j\frac{9 e^2}{4\pi\epsilon_0\vert z_i-z_j\vert^5}$ and
for $i\neq j$ we have $\gamma_{ij}=\frac{e^2}{2m\pi\epsilon_0
\vert z_i-z_j\vert^3}$, $\alpha_{ij}=\frac{9 e^2}{16\pi\epsilon_0
\vert z_i-z_j \vert ^5}$, $\kappa_{ij}=-\frac{3 e^2}
{8\pi\epsilon_0\vert z_i-z_j \vert^5}$, the coupling to axial
modes which exists in third and fourth
order has been omitted. While for $\omega_x \gg \omega_c$ the
contribution of the non-linear terms is small, this lowest order
expansion of the Hamiltonian shows that by lowering $\omega_x$ we
decrease $\gamma_{ii}$ leading to an increase of $\langle {\hat x}^2
\rangle$ and thus of the contribution of the nonlinear terms.
Eventually, for $\omega_x = \omega_c$ this will lead to a phase
transition between a linear chain and a zig-zag configuration.
Further decreasing $\omega_x$ results in more complex spatial
configurations \cite{Birkl KW 92}.

{\em Scalar quantum field with an adjustable mass --- } The
Hamiltonian (\ref{Hamiltonexpanded}) describes a non-linear scalar
field theory on a lattice whose effective mass may be adjusted via
the free parameter $\omega_x$\cite{correction}.
At the point of phase transition between the linear chain and the
zig-zag formation (fig.\ref{trans}) the field becomes massless
corresponding to a critical field. The behavior of the entropy of
single sites as well as its scaling with the size of blocks of sites
for critical and non-critical lattice field theories has been of
recent interest. While for non-critical systems an area law holds
\cite{Plenio EDC 05}, the block entanglement diverges at the
critical point at least in Gaussian 1-D models \cite{Cramer EP 07}.
Here we explore these questions numerically for the full non-linear
Hamiltonian (\ref{Hamiltonexpanded}) that may be realized in ion
traps and, in order to gain intuition, for a version of it that is
linearized about the equilibrium positions in the linear and the
zig-zag cases.

We begin with the linearized model for a chain of ions, i.e. the
regime where $\omega_x \gg \omega_c$,  and expand ${\hat x}_i$ and
${\hat p}_i$ for the radial motion at site $i$ into normal modes
of frequency $\omega_n$ with annihilation and creation operators
$a_n,a_n^{\dagger}$. Here ${\hat x}_i = \sum_n b_n^i \left({\hat
a}_n e^{- i w_n t} + {\hat a}_n^\dagger e^{i w_n
t}\right)\sqrt{\hbar/2 m w_n}$ and ${\hat p}_i = -i \sum_n b_n^i
\sqrt{\hbar m w_n/2}\left({\hat a}_n e^{- i w_n t} - {\hat
a}_n^\dagger e^{i w_n t}\right),$ where $b_n^i$, are the normal
mode coefficients of the crystal.   Then $ \langle x_i^2\rangle =
\frac{1}{2}\sum_n (b_n^i)^2 \hbar/m w_n$ and $\langle
p_i^2\rangle=\frac{1}{2}\sum_n (b_n^i)^2 \hbar m w_n.$ At the phase
transition between the linear and the zig-zag configuration one normal mode
becomes massless, i.e. its frequency $\omega_n$ vanishes. In a
linearized model the massless mode leads to diverging $\langle {\hat
x}^2 \rangle$. Since $\langle {\hat p}^2 \rangle$ remains finite the
entropy, which is a function of $\langle {\hat x}^2 \rangle \cdot
\langle {\hat p}^2 \rangle$, diverges \cite{eisert2003ijqi}. Only
sites that have vanishing amplitude in the zero mode would have a
finite $\langle x^2 \rangle$.

For $\omega_x<\omega_c$, the system posseses two nearly degenerate
ground states and a double well potential for a large crystal is
realized (Fig. \ref{trans}). Here the restriction to a 2-dimensional
setting is essential as otherwise the potential has a `Mexican hat'
structure allowing for continuous rotations rather than forming a
double well potential. For $\omega_x\ll\omega_c$ we then expect that
the system will possess slightly more than one bit of entropy per
site arising from the binary choice of location for each ion. The
excess is due to the entropy available to the ion within each
potential well.

At the transition point the entropy of a single ion of the harmonic
chain diverges logarithmically. The entropy of a harmonic
oscillator near the transition point satisfies $S_1(\omega_x) =
\log(\sqrt{\langle x^2 \rangle\langle p^2 \rangle}/\hbar) \approx
\log(|b_n^i|\sqrt{\langle p^2 \rangle/(2 \hbar m \omega_0)})$. Since,  the lowest normal mode frequency,
$\omega_0^2=\omega_x^2-\omega_c^2$, we find
\be
        S_1(\omega_x) \approx -\frac{1}{4}\log
        (\omega_x^2/\omega_z^2-\omega_c^2/\omega_z^2)=-\frac{1}{2}\log\omega_0/\omega_z.\label{entrel}
\ee For a small number of ions in the linearized model analytical
results for the entropy may be derived. For $\omega_x>\omega_c$ and two
ions the entropy is $S_2 = -\frac{1}{4} \log
(\omega_x^2/\omega_z^2-1) + \log \left(\frac{1}{4\sqrt{2}}\right) +
1 + O\left(\sqrt[4]{\omega_x^2/\omega_z^2-1}\right)$ while for the
middle ion in a three ion chain $S_3 = -\frac{1}{4} \log
\left(\omega_x^2/\omega_z^2-\frac{12}{5}\right)+\log
\left(\sqrt{2/5}/3\right)+1+O\left((\omega_x^2/\omega_z^2-12/5)^{1/4}\right).$

\begin{figure}
\begin{center}
\psfrag{n}{$\omega_x/\omega_z$}\psfrag{pop}{\tiny
Population}\psfrag{entropy}{Entropy}\psfrag{x}{\tiny Distance}
\includegraphics[width=0.50\textwidth,height=0.2\textheight]{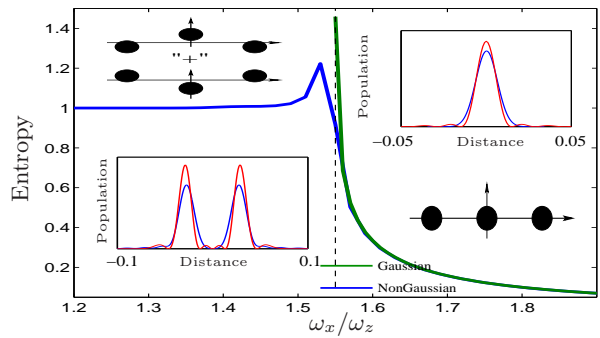}\\
\end{center}
\caption{Transition of three ions. The green curve shows the
entropy of the first ion as a function of $\omega_x$. At the
transition point $\omega_c/\omega_z=(12/5)^{1/2}$ the entropy
in the Gaussian approximation diverges and the real value (with
no approximations) reaches a maximum value after the transition
point. The insets show the probability distribution before (right)
and after (Left) the transition. The full numerics is in red and
the results obtained from considering only the normal mode are
in blue. The distance is in $4.5 nm$ units. \label{EntropyThree} }
\end{figure}

Following the discussion of the linearized models, we now analyze
numerically the phase transition between the linear chain and the
zig-zag configuration for the non-linear Hamiltonian
(\ref{Hamiltonexpanded}) for three ions. The transition between the
two phases is a second order transition\cite{Fishman2007} where the
order parameter is the displacement of the equilibrium position from
the axis as indicated by the Landau theory. A full numerical
calculation using Hamiltonian (\ref{Hamiltonexpanded}) without any
externally applied non-linearities is shown in
Fig.\ref{EntropyThree}. Far from the transition point
$\omega_x\gg\omega_c$ there is excellent agreement between the full
numerical calculation and the linearized models discussed above.
When approaching the transition the linearized models predict
diverging single site entropies while the full Hamiltonian
(\ref{Hamiltonexpanded}) yields a finite maximum value and then, for
$\omega_x\ll\omega_c$, the entropy approaches unity i.e. an EPR state between one
ion and the rest. The inset shows the comparison between the full
numerics and the results of treating the low normal mode as
decoupled, i.e., the red line results from the full numerics for
three ions and the blue line is obtained by considering only the
zero mode and then convolving the other modes to get the $x_1$
distribution. In order to compare the results we have checked the
equation $\langle x_1^2 \rangle = \frac{1}{6}(\langle x_0^2 \rangle
+ 2\langle x_{cm}^2 \rangle +3 \langle x_{b}^2 \rangle  )$, i.e.,
the $x_1$ deviation as a function of the normal mode deviations. The
correspondence is perfect except at the transition point where
$\sqrt{\langle x_1^2 \rangle}$ is $32 nm$ and the decoupled normal
mode result gives $38 nm$. For three ions the coefficient for the
quartic term is $b = 3\cdot 10^{-4} J\cdot m^{-4}$ which limits the
energy gap to $\approx 50kHz$. As can be seen from Fig.
\ref{EntropyThree} the nonlinear coupling between the different
normal modes is negligible. As the number of ions increases the
distances between the ions decreases; this raises the minimal energy
gap and reduces the maximal entanglement. In this region the chain becomes highly non
linear; this enables the observation of features peculiar to
nonlinear systems like solitons and quantum and classical discrete
breathers\cite{Fleurov}.

After this discussion of the critical entanglement
properties at the phase transition between the linear
chain and the zig-zag formation we now move to the
discussion of the second main feature of this system.

{\em Double well potential -- } The formation of double well
potentials in ion traps through the application of external
potentials has been considered in the past as an interesting system
for the exploration of the classical-quantum boundary and as a
system in which to preform atom interferometry. A number of groups are
currently working towards scaling up ion trap quantum information
processing to larger numbers of qubits following the general
approach suggested by \cite{Kielpinski2002}. A key step in this
approach is to separate two ions that are initially close together in a
particular trapping zone of an array of miniature traps, so that
they end up in separate trapping zones. This is done by raising the
potential on a small `separation' electrode. This electrode needs to
be very small in order to allow for the production of a steep
gradient in the potential between the ions. During the separation
protocol the trap transiently moves through a double well
configuration. The separation of a pair of ions in this way has been
reported using a separation electrode with an axial dimension of
100$\mu$m \cite{Rowe2002}. A new generation of traps with smaller
dimensions is currently under development worldwide but problems
with anomalously high heating rates in these very small traps have
yet to be solved. A theoretical study of relevance to this problem
indicates that higher field gradients may be generated with
relatively large electrodes that provide a dc octupole moment along
with the usual a.c. radial quadrupole \cite{Home2006}. Traps based
on these ideas have yet to be tested. Despite these advances in the
experimental ability to realize double well potentials for ions with a
measurable tunneling rate is still not within reach. However, as we
have explained above, the zig-zag configuration gives rise to
degenerate ground states separated by a tunnel barrier, thus
providing a natural double well potential requiring no externally
applied fields. As only the harmonic trapping potential
needs to be controlled, precise control of the depth and separation
of the effective double well potential and measurable tunneling rates
should be achievable. This moves the realization of double well
experiments in ion crystals within the reach of current experimental
technology.

We will now discuss various possibilities to control the parameters
of the double well, to rotate the two-level system formed by the
double well potential across the entire Bloch sphere and to measure
coherence and tunneling rates in the double well potential. In the
vicinity of the transition point the low normal mode alone controls
the transition. Therefore by controlling the parameters of the
Hamiltonian (\ref{Hamiltonexpanded}) and thus the values of $a$ and
$b$ in Eq.\ref{landau}, a double well structure with a distance
between the two minima is $2 (-a/b)^{1/2}$ and the width of the wave
function of $(\hbar/(2 \sqrt{- m a}))^{1/2}$ is created. The double
well is created at the stage when each well is deep enough to
accommodate one level, i.e., $\frac{1}{4} \frac{a^2}{b}\approx\hbar
\sqrt{2 a}$(the depth is of the order of one excitation).
Which means  $a \gtrsim 2^{5/3}(\hbar^2 b^2/m)^{1/3}$. For the zig-zag configuration for
three to ten ions $\sqrt{\omega_c^2-\omega_x^2}$ varies between
$76kHz$ to $155kHz$ and for the square configuration it is $36
kHz$.

Such double wells can be created for the zig-zag configurations. In
order to make the transition between the single well and the double
well adiabatic, the transition rate should be slower than the
minimal energy gap. The energy gap that should be considered is of
the order of the energy gap of the quartic potential. The ground state
in the Gaussian approximation is
$\frac{3^{4/3}}{2^{8/3}}\hbar^{4/3}b^{1/3}$ and the energy gap is
approximately $\left(\frac{3}{2}\right)^{4/3}\hbar^{4/3}b^{1/3}$. In the zig
zag configuration the minimal energy gap varies from $80kHz$
for three ions to $200kHz$ for seven ions. High energy gaps
increase the robustness of the system to decoherence at the
transition point.

\begin{figure}
\begin{center}
\psfrag{a}{$\omega_x=\omega_c$}
\includegraphics[width=0.40\textwidth,height=0.15\textheight]{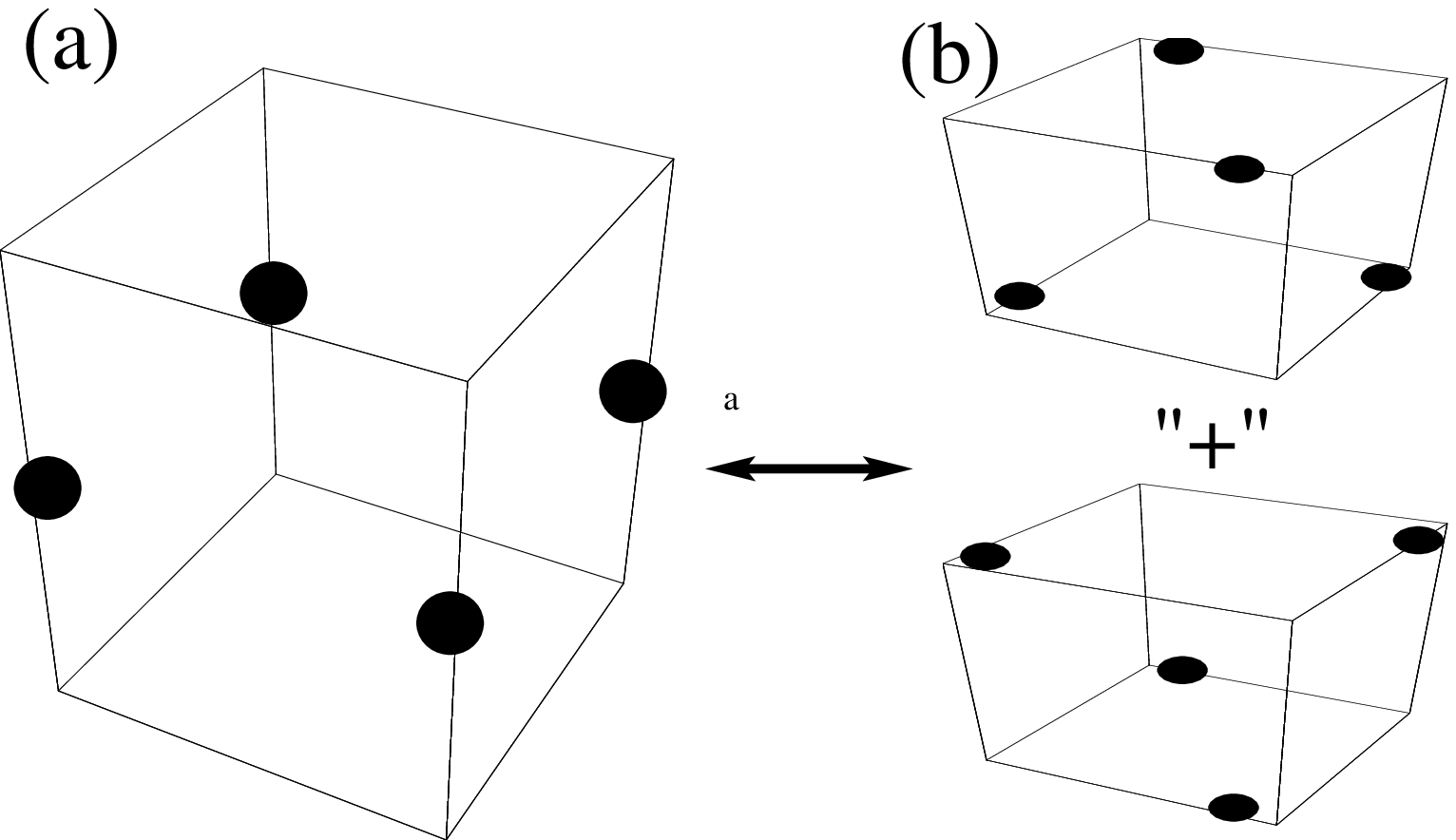}\\
\includegraphics[width=0.20\textwidth,height=0.15\textheight]{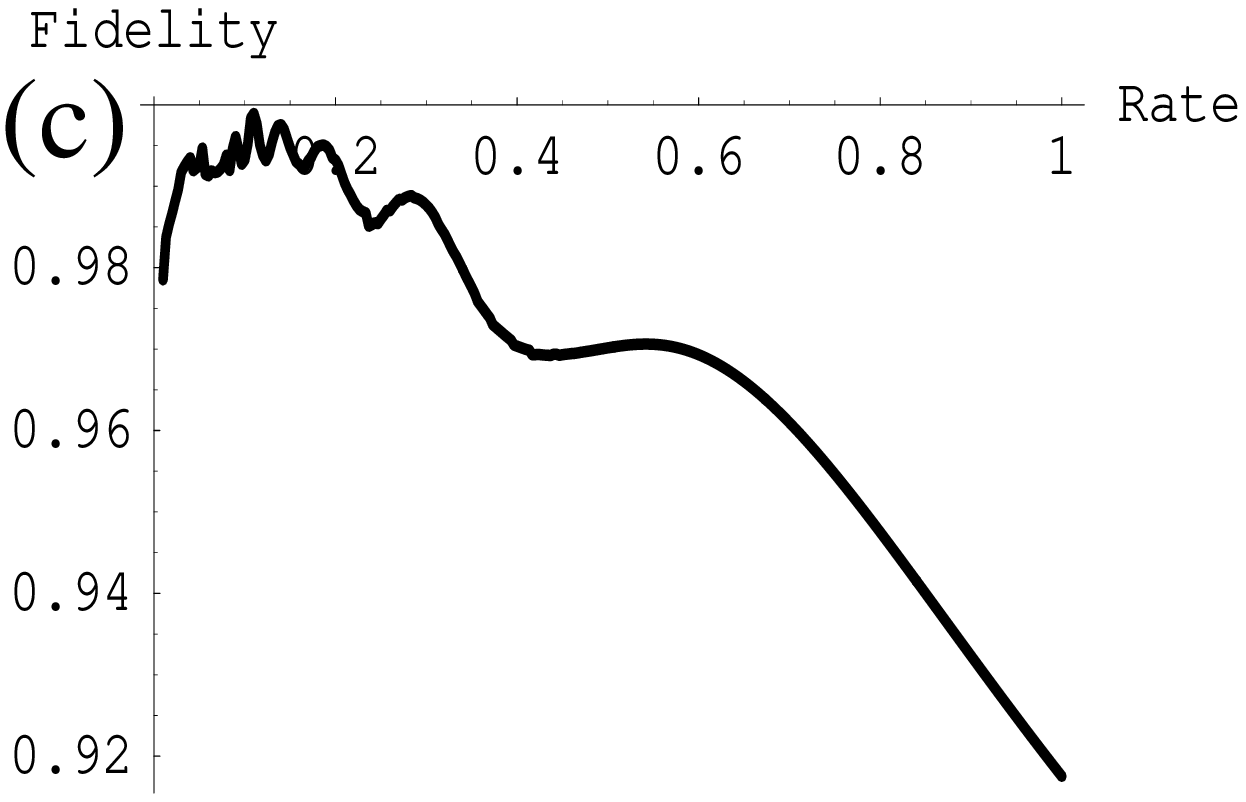}
\includegraphics[width=0.20\textwidth,height=0.15\textheight]{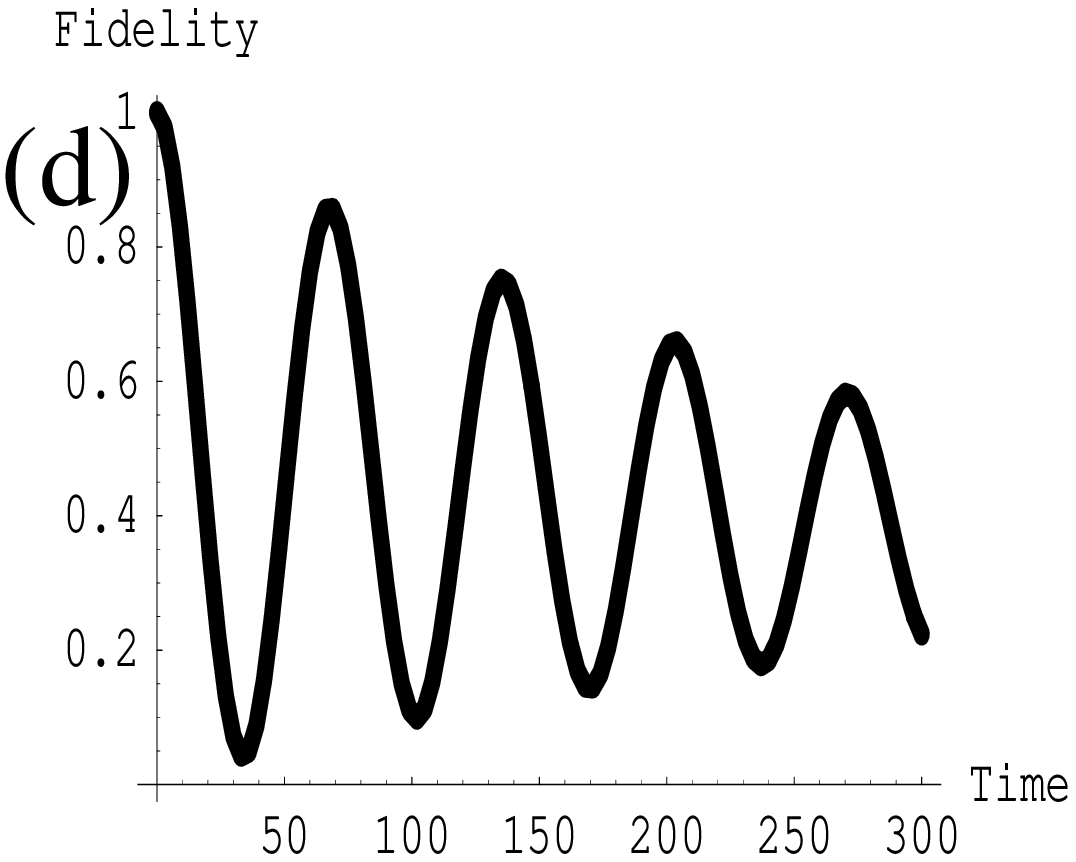}
\end{center}
\caption{(a) 1D Square configuration of ions. (b) is a  tetrahedron
phase. (c) From single well to double well adiabatic passage. Final
fidelity as a function of rate. In order to correct this error,
cooling in the double well phase could be possible. In this phase the
system would choose randomly the left or the right state. (d) Rabi
flipping for four ions}\label{Four}
\end{figure}

Double wells can also be realized in the transition between $2D$ and
$3D$. This transition may be more suitable for Penning trap crystals
which are not heated by micromotion. The  zig zag transition could be
realized both in Paul traps and in Penning traps after axialization.
In the case of four ions, for $\omega_x/\omega_z<0.822$ the stable
configuration is two dimensional in the $x-y$ plane
(Fig.\ref{Four}(a)). For $0.822<\omega_x/ \omega_z<1.27$ a
tetrahedron shown in Fig.\ref{Four}(b) is the stable configuration.
In the square configuration $b=5\cdot10^{-4}J\cdot m^{-4}$
thus the energy gap is $36kHz$ and the local potential is
$140kHz$ at the optimal double well point.

The
tunneling rates and the minimal energy gap increase with the number
of ions. For three ions in the zig zag configuration the optimal
point is at $\omega_x=\omega_c-3kHz$, where distance between
the configurations is $180 nm$ and the tunneling rate is
$26.7kHz$. For seven ions the optimal point is at
$\omega_x=\omega_c-5.7kHz$ the distance between the
configurations is $\approx 1\mu m$ and the tunneling rate is $55kHz$. For the square configuration the optimal point is at $\omega_x=\omega_z-1kHz$ and the distance is $\approx 300 nm$.
For the zig zag configuration the tunneling rates vary from
$3kHz$ for three ions to $4.5kHz$ for seven ions
and $1kHz$ for the square configuration Fig.\ref{Four}(a). The distance at
the optimal point varies between $30nm$ to $40nm$ and $20nm$ for
four ions.

\begin{figure}
\begin{center}
\psfrag{a}{$\ket{L+R}$}\psfrag{b}{$\ket{L-R}$}\psfrag{d}{$\ket{L}$}\psfrag{c}{$\ket{R}$}
\psfrag{khz}{$70kHz$}\psfrag{e}{$\ket{C}$}\psfrag{down}{$\ket{\down}$}\psfrag{up}{$\ket{\up}$}
\psfrag{delta}{$\delta$}\psfrag{w1}{$\omega_1$}\psfrag{w2}{$\omega_2$}\psfrag{W2}{$\Omega_2$}
\psfrag{W1}{$\Omega_1$}
\includegraphics[width=0.40\textwidth,height=0.15\textheight]{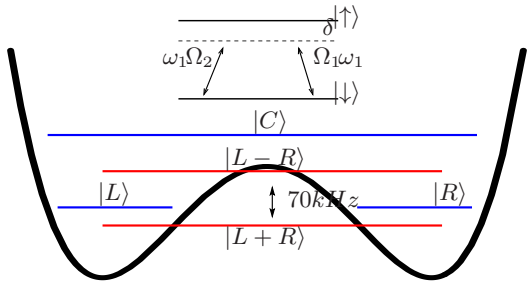}\\
\end{center}
\caption{The Double well energy levels for the three ions
case.}\label{double}
\end{figure}

Having created a double well potential in the way described above
the manipulation of its quantum state may be achieved in two ways.
The first approach maps the external degree of freedom into the
electronic one employing a Hamiltonian $\eta\Omega_1 (\sigma_+^{atom}
\ket{L-R}\bra{L+R} + \sigma_-^{atom}\ket{L+R}\bra{L-R} )$. This
is achieved employing a laser driving an electronic transition
with a detuning equalling the energy gap between $\ket {L-R}$
and $\ket {L+R}$ and is possible since in the regime where each
well supports one localized state and the spatial separation is less
than a $\mu m$, one laser can make the transitions as long as the
operation rate is slower than the energy gap fig.\ref{double}
($70kHz$ for three ions). Thus the manipulation of the external
degree of freedom is achieved by mapping it onto the electronic
degree of freedom, followed by its manipulation employing an
on-resonance laser and the subsequent mapping back onto the
external degree of freedom.

\begin{figure}
\begin{center}
\includegraphics[width=0.45\textwidth,height=0.25\textheight]{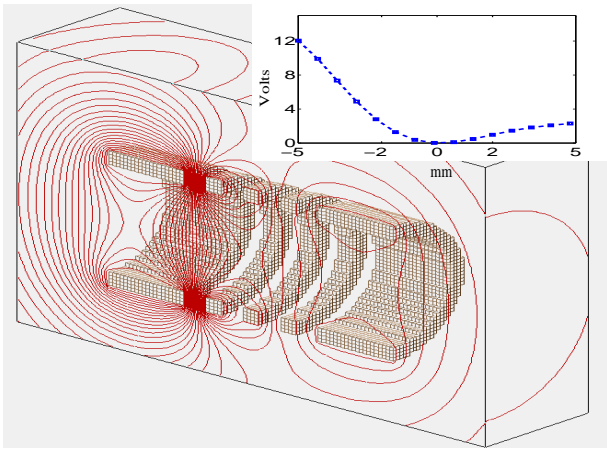}\\
\end{center}
\caption{Cut-away view of a cylindrical Penning trap consisting of 5 electrodes showing equipotentials calculated using SIMION. With the electrodes labelled a-e from left to right the applied voltages are: $V_a=15.8V$, $V_b=-3V$, $V_c=0V$, $V_d=3V$ and $V_d=4.5 V$. The inner diameter of the electrodes is $6mm$. The widths of the three inner electrodes (in the axial direction) are $0.8mm$ and the voltages applied to them have been chosen to generate a significant axial cubic term. The voltages on the outer electrodes are chosen to push the potential minimum back to the middle of the structure and to provide the usual quadratic term. The inset shows the axial potential in the central region of the trap.}
\label{cubic}
\end{figure}

For larger separation two lasers could create Raman
transitions between the $\ket L$ and $\ket R$(the ground states
of the the wells) states using an
excited state $|C\rangle$ of the system whose wave function extends
across both wells and whose energy is thus well above the gap
between the two lowest lying states of the double well potential.
The Raman transition could be created using the internal degrees of
freedom of the ion, by using the three state $\ket{\down
L}\longleftrightarrow \ket{\up C} \longleftrightarrow \ket {\down
R}$. Raman transitions for double well systems were suggested in
\cite{Pazy2001,Dum1998,Selsto2006}. For this purpose an asymmetry
between the two potential wells should be introduced in order to
distinguish between the right and the left well
which will break the symmetry between left and right. In order for
the cubic term($\alpha x^3$) to create an energy gap of $1kHz$,
$\alpha$ should be of the order of $10^{-10} \frac{J}{m^3}$. This will require application of $10V$ voltage for electrode size of $1 mm$. A cubic term can be created using a trap geometry shown in Fig. \ref{cubic}.

The manipulation of the external degree of freedom may also
be achieved by applying a radio frequency drive, i.e., by
applying a time varying potential $V_0 \cos \omega t$. This
will create the  effective $\sigma_x$ or $\sigma_y$ rotations.
Another way to create only one of these rotations is by adding
a small cubic interaction which will add a phase shift between
the $\ket L$ and $\ket R$ states.

An alternative approach to the manipulation of the state of the
double well system exploits the fact that we are able to control the
state of a single well potential.
Thus by moving back and forth across the
phase transition between single well and double well potential
adiabatically we can generate any desired state in the double well
potential by controlling the state in the single well potential. The
state $\ket 0$ of the single well would change adiabatically to
$\ket L - \ket R$ in the double well while the state $\ket 1 $
evolves into $\ket L + \ket R$. Due to the nonlinear term - $b
x^4$, the minimal gap would not vanish in this process and thus the
transition can be carried out at a sufficiently high rate while
remaining approximately adiabatic. Thus, in order to move
adiabatically from one phase to the other, the rate should be less
than the minimal energy gap but higher than the decoherence rate (Fig.\ref{Four}(c)).
Arbitrary measurements on the double well may be achieved by combining
unitary rotations implemented as explained above with measurements
of the presence or absence of the system in the left or right well.

In the following we show a specific example of the transition
between the two phases. For four ions, in the transition between a
2D and a 3D structure, the adiabatic condition is restrictive at the
transition point $\omega_0=0$, at which the potential is
$V=\frac{1}{4} b x^4$. For this transition the the minimal energy
gap between the ground state and the first excited state is
$50kHz$. This implies that the adiabatic sweeping could be
relatively fast and create a small number of excitations. Fig.
\ref{Four} shows the overlap of the state at a specific time
of the transition, with the instantaneous ground state. The
faster the rate the smaller the overlap. Fig. \ref{Four}(d)
shows the Rabi flipping for four ions between the different
states in Fig. \ref{Four}(b). It can be seen that a few cycles
could be measured. Another way to measure the state is to
change the system adiabatically to the state where the distance
between the two wells is larger than one wavelength and then
to measure, the $\ket L$ and $\ket R$ states using lasers.

The realization of the double well described here could open the way to
interferometry at the Nano-scale. Ion traps have already been used as a
measurement apparatus at the nanoscale \cite{Gutho}. Double wells
for single atoms and BEC's have been used for precise measurement in
interferometry experiments. Using the Ion trap double well,
measurements of non linear electric fields on the nanoscale should be
possible. Measurements via interferometry should be more precise than
the single well measurements. Moreover since linear fields are not
seen by the ion trap, this procedure could measure cubic fields
while completely eliminating the linear contribution.
This double well could also measure magnetic field gradients.
The initial state of the internal degrees of freedom can be initialized in a superposition of
two levels and the phase induced by the magnetic gradients can be measured by the population of the excited state.

The ideas presented here may be realized in linear Paul traps
as well as Penning traps. Linear ion traps will exhibit
micro-motion when ions are displaced from the trap axis.
Nevertheless, for small deviations as discussed here, micromotion
and the resulting decoherence effects will not be significant.
The advantage of the Penning trap is that it does not suffer
from micro-motion, and thus the crystal may be less fragile
to decoherence. However, observations need to take place
in a rotating frame in which the effect of the magnetic
field is cancelled \cite{Thompson1997}.

\section{Discussion}
We have suggested a realization of a non-linear Klein-Gordon
field on a lattice with adjustable parameters. For a small
number of ions the crystal may be cooled to the ground
state and entanglement properties may be measured to a high
precision. We have demonstrated that by crossing the quantum
phase transition between a linear and a zig-zag configuration
a fully controllable double well potential may be achieved
with available technology. This demonstrates the potential
of this system for the study of complex quantum many body
phenomena. By controlling the nonlinear terms for example,
highly non - linear excitations, such as kinks and breathers,
can be created. Furthermore, rapidly driving the system
through the phase transition, the Kibble mechanism may be
observed \cite{forthcoming}, due to the ability of of
addressing and measuring individual ions with high precision.

\section{Acknowledgments}

We acknwoledge support by the European Commission under the
Integrated Project QAP, the Royal Society, the
EPSRC QIP-IRC and EPSRC grant number EP/E045049. Helpful
discussion with M. Hartmann and F.G.S.L Brand{\~a}o are
acknowledged.

\bibliography{Master}

\begin{thebibliography}{34}
\expandafter\ifx\csname
natexlab\endcsname\relax\def\natexlab#1{#1}\fi
\expandafter\ifx\csname bibnamefont\endcsname\relax
  \def\bibnamefont#1{#1}\fi
\expandafter\ifx\csname bibfnamefont\endcsname\relax
  \def\bibfnamefont#1{#1}\fi
\expandafter\ifx\csname citenamefont\endcsname\relax
  \def\citenamefont#1{#1}\fi
\expandafter\ifx\csname url\endcsname\relax
  \def\url#1{\texttt{#1}}\fi
\expandafter\ifx\csname urlprefix\endcsname\relax\def\urlprefix{URL
}\fi \providecommand{\bibinfo}[2]{#2}
\providecommand{\eprint}[2][]{\url{#2}}

\bibitem[{\citenamefont{Hund}(1927)}]{hund1927}
\bibinfo{author}{\bibfnamefont{F.}~\bibnamefont{Hund}}, \bibinfo{journal}{Z.
  Phys.} \textbf{\bibinfo{volume}{43}}, \bibinfo{pages}{803}
  (\bibinfo{year}{1927}).

\bibitem{LandauTheory} L. D. Landau and E. M. Lifshitz, {\it Statistical
Physics} (Pergamon, Oxford, 1958).

\bibitem{Osterloh 02} A. Osterloh, L. Amico, G. Falci, R. Fazio,
Nature {\bf 416}, 608 (2002).

\bibitem{Jin 04} B.-Q. Jin and V.E. Korepin, J. Stat. Phys.
{\bf 116}, 79 (2004).

\bibitem{Audenaert EPW 02} K. Audenaert, J. Eisert, M.B. Plenio
and R.F. Werner, Phys. Rev. A {\bf 66}, 042307 (2002);
A. Botero and B. Reznik, Phys. Rev. A {\bf 67}, 052311 (2003).

\bibitem{Cramer EP 07} M. Cramer, J. Eisert and M.B. Plenio,
Phys. Rev. Lett. {\bf 98}, 220603 (2007).

\bibitem[{\citenamefont{Leibfried et~al.}(1987)\citenamefont{Leibfried,
  DeMarco, Meyer, Lucas, Barrett, Britton, Itano, and
  Jelenkovi}}]{Leibfried2003}
\bibinfo{author}{\bibfnamefont{D.}~\bibnamefont{Leibfried}},
  \bibinfo{author}{\bibfnamefont{B.}~\bibnamefont{DeMarco}},
  \bibinfo{author}{\bibfnamefont{V.}~\bibnamefont{Meyer}},
  \bibinfo{author}{\bibfnamefont{D.}~\bibnamefont{Lucas}},
  \bibinfo{author}{\bibfnamefont{M.}~\bibnamefont{Barrett}},
  \bibinfo{author}{\bibfnamefont{J.}~\bibnamefont{Britton}},
  \bibinfo{author}{\bibfnamefont{W.~M.} \bibnamefont{Itano}}, \bibnamefont{and}
  \bibinfo{author}{\bibfnamefont{B.}~\bibnamefont{Jelenkovi}},
  \bibinfo{journal}{Nature} \textbf{\bibinfo{volume}{422}},
  \bibinfo{pages}{412} (\bibinfo{year}{1987}).

\bibitem[{\citenamefont{Retzker et~al.}(2005)\citenamefont{Retzker, Cirac, and
  Reznik}}]{retzker2005}
\bibinfo{author}{\bibfnamefont{A.}~\bibnamefont{Retzker}},
  \bibinfo{author}{\bibfnamefont{I.~J.} \bibnamefont{Cirac}}, \bibnamefont{and}
  \bibinfo{author}{\bibfnamefont{B.}~\bibnamefont{Reznik}},
  \bibinfo{journal}{Phys. Rev. Lett.} \textbf{\bibinfo{volume}{94}},
  \bibinfo{pages}{050504} (\bibinfo{year}{2005}).

\bibitem[{\citenamefont{Alsing et~al.}(2005)\citenamefont{Alsing, Dowling, and
  Milburn}}]{Alsing2005}
\bibinfo{author}{\bibfnamefont{P.}~\bibnamefont{Alsing}},
  \bibinfo{author}{\bibfnamefont{J.}~\bibnamefont{Dowling}}, \bibnamefont{and}
  \bibinfo{author}{\bibfnamefont{G.}~\bibnamefont{Milburn}},
  \bibinfo{journal}{Phys. Rev. Lett.} \textbf{\bibinfo{volume}{94}},
  \bibinfo{pages}{220401} (\bibinfo{year}{2005}).

\bibitem{Porras C 04} D. Porras and J.I. Cirac,
Phys. Rev. Lett. {\bf 92}, 207901 (2004). 

\bibitem[{\citenamefont{Deng et~al.}(2005)\citenamefont{Deng, Porras, and
  Cirac}}]{Porras2005a}
\bibinfo{author}{\bibfnamefont{X.-L.} \bibnamefont{Deng}},
  \bibinfo{author}{\bibfnamefont{D.}~\bibnamefont{Porras}}, \bibnamefont{and}
  \bibinfo{author}{\bibfnamefont{J.~I.} \bibnamefont{Cirac}},
  \bibinfo{journal}{Phys. Rev. A} \textbf{\bibinfo{volume}{72}},
  \bibinfo{pages}{063407} (\bibinfo{year}{2005}).

\bibitem[{\citenamefont{Porras and Cirac}(2004)}]{Porras2004a}
\bibinfo{author}{\bibfnamefont{D.}~\bibnamefont{Porras}} \bibnamefont{and}
  \bibinfo{author}{\bibfnamefont{J.~I.} \bibnamefont{Cirac}},
  \bibinfo{journal}{Phys. Rev. Lett.} \textbf{\bibinfo{volume}{93}},
  \bibinfo{pages}{263602} (\bibinfo{year}{2004}).

\bibitem[{\citenamefont{Dubin and O\char39{}Neil}(1999)}]{Dubin1999}
\bibinfo{author}{\bibfnamefont{D.~H.~E.} \bibnamefont{Dubin}} \bibnamefont{and}
  \bibinfo{author}{\bibfnamefont{T.~M.} \bibnamefont{O\char39{}Neil}},
  \bibinfo{journal}{Rev. Mod. Phys.} \textbf{\bibinfo{volume}{71}},
  \bibinfo{pages}{87} (\bibinfo{year}{1999}).

\bibitem[{\citenamefont{Rafac et~al.}(1991)\citenamefont{Rafac, Schiffer,
  Hangst, Dubin, and Wales}}]{Rafac1991}
\bibinfo{author}{\bibfnamefont{R.}~\bibnamefont{Rafac}},
  \bibinfo{author}{\bibfnamefont{J.}~\bibnamefont{Schiffer}},
  \bibinfo{author}{\bibfnamefont{J.}~\bibnamefont{Hangst}},
  \bibinfo{author}{\bibfnamefont{D.}~\bibnamefont{Dubin}}, \bibnamefont{and}
  \bibinfo{author}{\bibfnamefont{D.}~\bibnamefont{Wales}},
  \bibinfo{journal}{Proc. Natl. Acad. Sci} \textbf{\bibinfo{volume}{88}},
  \bibinfo{pages}{483} (\bibinfo{year}{1991}).

\bibitem[{\citenamefont{Wineland et~al.}(1987)\citenamefont{Wineland,
  Bergquist, Itano, Bollinger, and Manney}}]{Wineland1987}
\bibinfo{author}{\bibfnamefont{D.~J.} \bibnamefont{Wineland}},
  \bibinfo{author}{\bibfnamefont{J.~C.} \bibnamefont{Bergquist}},
  \bibinfo{author}{\bibfnamefont{W.~M.} \bibnamefont{Itano}},
  \bibinfo{author}{\bibfnamefont{J.~J.} \bibnamefont{Bollinger}},
  \bibnamefont{and} \bibinfo{author}{\bibfnamefont{C.~H.}
  \bibnamefont{Manney}}, \bibinfo{journal}{Phys. Rev. Lett.}
  \textbf{\bibinfo{volume}{59}}, \bibinfo{pages}{2935} (\bibinfo{year}{1987}).

\bibitem{Birkl KW 92} G. Birkl, S. Kassner and H. Walther,
Nature {\bf 357}, 310 (1992).

\bibitem[{\citenamefont{Casdorff and Blatt}(1998)}]{Casdorff1998}
\bibinfo{author}{\bibfnamefont{R.}~\bibnamefont{Casdorff}} \bibnamefont{and}
  \bibinfo{author}{\bibfnamefont{R.}~\bibnamefont{Blatt}},
  \bibinfo{journal}{App. Phys. B} \textbf{\bibinfo{volume}{45}},
  \bibinfo{pages}{175} (\bibinfo{year}{1998}).

\bibitem{Morigi F 04} G. Morigi and S. Fishman, Phys. Rev. E {\bf 70},
066141 (2004).

\bibitem{axialization}
Breaking the radial symmetry in a linear RF trap can be achieved
straightforwardly by applying a DC potential to two opposing rods
leading to a deeper potential along the direction between these
rods. In the Penning trap this can be achieved through the
technique of axialisation \cite{Powell 02} where a radial
quadrupole potential that rotates at half the true cyclotron
frequency is applied. The effect of this potential is a static
distortion of the 2-D harmonic well in the rotating frame so
that the trap frequencies along two orthogonal axes in
this frame (say $x^\prime$ and $y^\prime$) are different.
For the application envisaged here one would apply a potential
in the region of a few volts.

\bibitem{nonlinear} The nonlinear part could be treated as second
order correction as long as the fluctuation of the zero mode are
smaller than the distance between the ions, or $\Delta x_0/ z_0 \ll
\omega_0/\omega_z $.

\bibitem{correction} Unlike in the regular scalar field on a
lattice, a long range coupling exist and the sign of the coupling
terms is positive.

\bibitem{Plenio EDC 05} M.B. Plenio,
J. Eisert, J. Dreissig, M. Cramer, Phys. Rev. Lett. {\bf 94},
060503 (2005); M. Cramer, J. Eisert, M.B. Plenio and J. Dreissig,
Phys. Rev. A {\bf 73}, 012309 (2006).

\bibitem{eisert2003ijqi} J. Eisert and M.B. Plenio,
Int. J. Quant. Inf. {\bf 1}, 479 (2003).

%

\bibitem[{\citenamefont{Thompson and Wilson}(1997)}]{Thompson1997}
\bibinfo{author}{\bibfnamefont{R.}~\bibnamefont{Thompson}} \bibnamefont{and}
  \bibinfo{author}{\bibfnamefont{D.}~\bibnamefont{Wilson}},
  \bibinfo{journal}{Zeits. Phys. D} \textbf{\bibinfo{volume}{42}},
  \bibinfo{pages}{271} (\bibinfo{year}{1997}).

\bibitem[{\citenamefont{Fishman et~al.}()\citenamefont{Fishman, Chiara,
  Calarco, and Morigi}}]{Fishman2007}
\bibinfo{author}{\bibfnamefont{S.}~\bibnamefont{Fishman}},
  \bibinfo{author}{\bibfnamefont{G.~D.} \bibnamefont{Chiara}},
  \bibinfo{author}{\bibfnamefont{T.}~\bibnamefont{Calarco}}, \bibnamefont{and}
  \bibinfo{author}{\bibfnamefont{G.}~\bibnamefont{Morigi}},
  \bibinfo{journal}{arXiv:0710.1831}  (2007).

\bibitem[{\citenamefont{V}(2003)}]{Fleurov}
\bibinfo{author}{\bibfnamefont{V. Fleurov}}, \bibinfo{journal}{CHAOS}
\textbf{\bibinfo{volume}{13}}, \bibinfo{pages}{676} (\bibinfo{year}{2003}).

\bibitem{Powell 02} H.F. Powell, D.M. Segal and R.C. Thompson,
Phys. Rev. Lett. {\bf 89}, 093003 (2002).


\bibitem{Kielpinski2002}
D. Kielpinski, C. Monroe and  D. J. Wineland, Nature  {\bf  417},
709( 2002).

\bibitem{Rowe2002} M.A. Rowe {\em et al},
Quant. Inf. Comp. {\bf 2}, 257 (2002).

\bibitem{Home2006} J.P. Home and A.M. Steane,  Quant. Inf. Comp.
{\bf 6}, 289 (2006)

\bibitem[{\citenamefont{Pazy et~al.}(2001)\citenamefont{Pazy, D\char39{}Amico,
  Zanardi, and Rossi}}]{Pazy2001}
\bibinfo{author}{\bibfnamefont{E.}~\bibnamefont{Pazy}},
  \bibinfo{author}{\bibfnamefont{I.}~\bibnamefont{D\char39{}Amico}},
  \bibinfo{author}{\bibfnamefont{P.}~\bibnamefont{Zanardi}}, \bibnamefont{and}
  \bibinfo{author}{\bibfnamefont{F.}~\bibnamefont{Rossi}},
  \bibinfo{journal}{Phys. Rev. B} \textbf{\bibinfo{volume}{64}},
  \bibinfo{pages}{195320} (\bibinfo{year}{2001}).

\bibitem[{\citenamefont{Dum et~al.}(1998)\citenamefont{Dum, Cirac, Lewenstein,
  and Zoller}}]{Dum1998}
\bibinfo{author}{\bibfnamefont{R.}~\bibnamefont{Dum}},
  \bibinfo{author}{\bibfnamefont{J.~I.} \bibnamefont{Cirac}},
  \bibinfo{author}{\bibfnamefont{M.}~\bibnamefont{Lewenstein}},
  \bibnamefont{and} \bibinfo{author}{\bibfnamefont{P.}~\bibnamefont{Zoller}},
  \bibinfo{journal}{Phys. Rev. Lett.} \textbf{\bibinfo{volume}{80}},
  \bibinfo{pages}{2972} (\bibinfo{year}{1998}).

\bibitem[{\citenamefont{Selsto and Forre}(2006)}]{Selsto2006}
\bibinfo{author}{\bibfnamefont{S.}~\bibnamefont{Selsto}} \bibnamefont{and}
  \bibinfo{author}{\bibfnamefont{M.}~\bibnamefont{Forre}},
  \bibinfo{journal}{Phys. Rev. B} \textbf{\bibinfo{volume}{71}},
  \bibinfo{pages}{195327} (\bibinfo{year}{2006}).

\bibitem{Gutho}  G. R. Guth\"ohrlein et al,  Nature  {\bf 414},
 49 (2001).

\bibitem{Snadden} M. J. Snadden et al, Phys. Rev. Lett. {\bf 81},
971(1998).

\bibitem{forthcoming} Work in progress

\end{thebibliography}
\end{document}